\documentclass[twocolumn,showpacs,prl]{revtex4-1}
\usepackage{amsmath,amssymb}
\usepackage{graphicx}
\usepackage{verbatim}

\draft
\begin{document}

\title{Correlation effects on topological insulator}

\author{Xiong-Jun Liu, Yang Liu and Xin Liu}
\affiliation{Department of Physics, Texas
A\&M University, College Station, Texas 77843-4242, USA}
%{b. Department of Physics, University of California, San Diego,
%California 92093, USA}

\begin{abstract}
The strong correlation effects on topological insulator are studied in a two-sublattice system with an onsite single-particle energy difference $\Delta$ between two sublattices. At $\Delta=0$, increasing the onsite interaction strength $U$ drives the transition from the quantum spin Hall insulating state to the non-topological antiferromagnetic Mott-insulating (AFMI) state. When $\Delta$ is larger than a certain value, a topologically trivial band insulator or AFMI at small values of $U$ may change into a quantum anomalous Hall state with antiferromagnetic ordering at intermediate values of $U$. Further increasing $U$ drives the system back into the topologically trivial state of AFMI. The corresponding phenomena is observable in the solid state and cold atom systems. We also propose a scheme to realize and detect these effects in cold atom systems.
\end{abstract}
\pacs{73.43.-f, %71.70.Ej,
71.30.+h, 05.30.Fk}
\date{\today}
\maketitle

%\baselineskip=16pt

%Topological phase transitions are an exciting topic in condensed matter physics which has attracted great attention since the discovery of the quantum Hall effect (QHE) \cite{QHE1}.

The recent prediction and experimental observation of the quantum spin Hall (QSH) state raise great interests in the study of topological phase transitions \cite{T1,T1',T1'',T2,T3}. A simple model for quantum spin Hall effect (QSHE) can be constructed by coupling the particles to a spin-dependent effective magnetic field \cite{Bernevig,zhu}. By extending such model to the many-body case, the possible appearance of fractional QSH states was studied \cite{Bernevig,liu2,Levin,Maciejk}. However, the more realistic situation for the QSH effect is the topological insulator (TI) which can be achieved with the inverted band structure in the presence of strong spin-orbit coupling. In the single particle picture the TI is characterized by the nontrivial $Z_2$ topological invariant, with one-dimensional gapless helical modes existing along the edge for two dimensional (2D) materials \cite{T1}, and (2+1)-dimensional ((2+1)D) gapless Dirac modes on the surface for the three dimensional (3D) case \cite{T2}. A natural generalization of the TI is to consider the many-body effect, of which the simplest way is to combine the TI and the Fermi-Hubbard model by considering the on-site interaction. Different correlation effects on the TI are supposed to be obtained by adjusting the interaction strength.

In this work we study the correlation effect on TI in a quasi 2D square lattice in the presence of a staggered sublattice potential $\Delta$, which is shown to have a pronounced effect on the topological properties of the antiferromagnetic Mott insulator (AFMI) obtained in the large on-site Hubbard interaction $U$. While in the case $\Delta<\Delta_c$, the AFMI is always topologically trivial, when $\Delta>\Delta_c$ it can be topologically non-trivial and exhibits the quantum anomalous Hall effect.

%In the case $\Delta<\Delta_c$, a non-topological antiferromagnetic Mott insulator (AFMI) phase is obtained at large particle-particle interaction $U$. However, it is remarkable that when $\Delta>\Delta_c$, we find the increasing $U$ may drive the system into a topologically non-trivial AFMI state which exhibits the quantum anomalous Hall effect (QAHE).

%
\begin{figure}[ht]
\includegraphics[width=1.0\columnwidth]{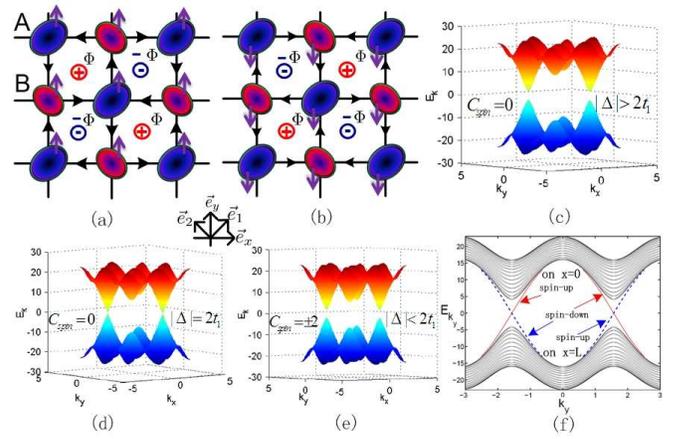}
\caption{(Color online) Configuration of 2D anisotropic square lattice for spin-up (a) and spin-down fermions (b); (c-e) Single particle bulk spectrum; (f) Single particle edge spectrum.} \label{fig1}
\end{figure}
We start with the quasi 2D square lattice model at half-filling depicted in
Fig. \ref{fig1}(a-b), with the on-site Hubbard interaction $U$ and a single-particle sublattice energy difference $\Delta$ between $A$ and $B$ sites. Similar to Haldane's original idea for QAHE \cite{Haldane} and the Kane-Mele (KM) model for QSHE \cite{T1}, we introduce a spin-dependent staggered magnetic flux in the lattice, which leads to a Peierls phase $\phi$ for the nearest-neighbor-site hopping along the marked direction (Fig. 1(a-b)). The Hamiltonian in the tight-binding form reads $H=H_0+H_{int}$, with
\begin{eqnarray}\label{eqn:tightbinding1}
H_0&=&-\sum_{<i,j>}t_0\bigr(\cos\phi+is_z\nu_{ij}\sin\phi\bigr)\hat c_{a,i}^\dag\hat c_{b,j}-\nonumber\\
&&-\sum_{\ll i,j\gg}\sum_{\substack{\mu=a,b\\l=1,2}}t_{\mu l}\hat c_{\mu,i}^\dag\hat c_{\mu,j}+\Delta\sum_{i}(n_{i,a}-n_{i,b}),
\nonumber\\
H_{int}&=&\sum_iUn_{i\uparrow}n_{i\downarrow},
\end{eqnarray}
where $c_{\mu}=(c_{\mu\uparrow},c_{\mu\downarrow})$, $\nu_{ij}=1$ ($-1$) for hopping along (opposite to) the marked direction, $t_0$ is the nearest-neighbor hopping coefficient and $U$ the intrasite interacting energy. In this model we consider the local orbitals on $A$ and $B$ sites to be anisotropic and thus the next-neighbor hopping (in $\vec e_{1,2}$ directions) coefficients $t_{al}\neq t_{bl}$ with $l=1,2$. Although the above system serves as a toy model, we stress later that the physics studied in the following is quite general. Furthermore, we shall propose a scheme to observe the phenomenon with cold atoms.

Before moving to the study of the correlation effect, we give a quick description of the topological phase transition in the absence of interaction. For convenience we transfer the original Hamiltonian into $\bold k$ space $H=\sum_{\bold k}\hat {\mathcal C}^{\dag}(\bold
k)\mathcal{H}_0(\bold k)\hat {\mathcal C}(\bold k)+H_{int}$ with $\hat
{\mathcal C}(\bold k)=(\hat c_{a\uparrow}(\bold k), \hat c_{b\uparrow}(\bold k),\hat c_{a\downarrow}(\bold k), \hat c_{b\downarrow}(\bold k))^T$ and obtain
\begin{eqnarray}\label{eqn:Hamiltonian1}
\mathcal{H}_0(\bold k)&=&\sum_{\alpha=1}^5d_{\alpha}(\bold k)\Gamma^\alpha,\nonumber\\
H_{int}&=&\frac{2U}{N_0}\sum_{\substack{\bold k\bold k'\bold q\\ \mu=a,b}}\hat c_{\mu,\bold k'+\bold q\uparrow}^\dag\hat c_{\mu,\bold k-\bold q\downarrow}^\dag\hat c_{\mu,\bold k+\bold q\downarrow}\hat c_{\mu,\bold k'-\bold q\uparrow},
\end{eqnarray}
where $d_1=-t_0\cos\phi(\cos k_xa+\cos k_ya), d_2=\Delta-2(t_{a1}+t_{a2}-t_{b1}-t_{b2})\cos k_xa\cos k_ya+2(t_{a1}+t_{b2}-t_{a2}-t_{b1})\sin k_xa\sin k_ya, d_3=d_4=0$, $d_5=-2t_0\sin\phi(\cos k_xa-\cos k_ya)$, $\Gamma^\alpha$ are defined via $\{\Gamma^{\alpha}\}=\{\sigma_x\otimes I,\sigma_z\otimes I,\sigma_y\otimes s_x,\sigma_y\otimes s_y,\sigma_y\otimes s_z\}$ and $N_0$ is the number of particles. For convenience we choose in the following $\phi=\pi/4, t_{a1}>t_{b1}, t_{a2}=t_{b2}$ and denote by $t_1=t_{a1}-t_{b1}$. We can then simplify the coefficient $d_2(\bold k)=\Delta+2t_1(\sin k_xa\sin k_ya-\cos k_xa\cos k_ya)$. Noting that $\Gamma^\alpha$ are even under time-reversal (TR) transformation and $d_{\alpha}(\bold k)=d_{\alpha}(-\bold k)$, thus the above Hamiltonian satisfies the TR symmetry.

The single particle spectrum of $\mathcal{H}_0(\bold k)$ is given in Fig. \ref{fig1}(c-e). When  $t_1=0$ and $\Delta>0$, the system is a \textit{trivial} band insulator. When $2t_1=\Delta$, the system becomes a semi-metal with the gap closed at $k_x=-k_y=\pi/2$ and $-\pi/2$ for spin-up and -down states, respectively. Around gap closing points the system is described by the massless Dirac Hamiltonians. Furthermore, when $2t_1>|\Delta|$, the gap opens again and the mass terms of the Dirac Hamiltonians on the two points change sign relative to those in the case $2t_1<|\Delta|$, which leads to a quantum jump ($\pm e^2/h$) of the Hall conductance for each Dirac Hamiltonian \cite{conductance}. Therefore the charge Hall conductivity (CHC) reads now $\sigma_{xy}^\uparrow=-\sigma_{xy}^\downarrow=e^2/h$ for spin-up and -down states. For this we know the spin Chern number of the present system: $C_{spin}=\pm2$ when $2t_1>|\Delta|$ and $C_{spin}=0$ otherwise, while the total CHC is always zero due to TR symmetry in the single-particle picture. The gapless helical edge modes on the boundaries at $x=0$ and $x=L$ are shown in Fig. \ref{fig1}(f), and such modes are protected by a nontrivial $Z_2$ topological number. In solid state systems, the edge modes can be detected by measuring the effective one-dimensional (1D) channel conductance \cite{T3}, and in cold atoms, these modes may be detected with light Bragg scattering \cite{liu3}.

Now we proceed to study the correlation effect on the topological phase transition. When $|U|$ is much smaller than the single-particle bulk gap, i.e. $|U|\ll2t_1-|\Delta|$, only the edge states will be scattered by particle-particle interaction, with the bulk states unaffected. In this way the effective Hamiltonian for edge modes can be described by 1D helical Luttinger liquid model with only the forward scattering, which cannot open a gap but leads to a renormalization of the group velocity of edge states according to the standard bosonization approach \cite{wu}.

It is more interesting to consider the strong interacting regime where $U$ is larger than the bulk gap. In this case the bulk physics will be changed and can be studied with Hartree-Fock mean field approach. For the half-filling system, the AF order (spin density wave (SDW)) will appear when the interacting strength $U$ exceeds some critical value. We can introduce the mean-field staggered AF order parameter with $m_i=\langle n_{i\uparrow}\rangle-\langle n_{i\downarrow}\rangle=\mbox{sgn}(i)m$,
where $\mbox{sgn}(i)=\pm1$ for A and B sub-lattices, respectively. Then the mean-field interacting Hamiltonian reads
\begin{eqnarray}\label{eqn:Hamiltonian3}
H_{int}^{m}=\frac{U}{4}\sum_i\langle n_i^2\rangle-\frac{U}{4}\sum_i\hat {\mathcal C}^{\dag}(\bold
k)\bigr[2m\Gamma^{15}-m_i^2\bigr]\hat {\mathcal C}(\bold k)
\end{eqnarray}
with $n_i=n_{i\uparrow}+n_{i\downarrow}$ and $\Gamma^{15}=s_z\otimes\sigma_z$. Note the inversion symmetry of the present system is broken when $\Delta\neq0$, for which the occupation numbers $\langle n_{ai}\rangle$ (at $A$ sites) and $\langle n_{bi}\rangle$ (at $B$ sites) are generally different. For convenience we denote by $\langle n_{ai}\rangle=1+\langle\delta n\rangle$ and $\langle n_{bi}\rangle=1-\langle\delta n\rangle$, where $\delta n$ characterizes the difference between the occupation numbers and should be determined self-consistently. By combination of Eqs. (\ref{eqn:Hamiltonian1}) and (\ref{eqn:Hamiltonian3}) we get now the Hamiltonian $H_{m}=\sum_{\bold k}\hat {\mathcal C}^{\dag}(\bold
k)\mathcal{H}_0^{(m)}\hat {\mathcal C}(\bold k)+\frac{N_0}{4}U(1+\langle\delta n\rangle^2+m^2)$,
where $\mathcal{H}_0^{(m)}(\bold k)=\mathcal{H}_0(\bold k)+d_{15}\Gamma^{15}$ with $d_{15}=mU/2$. Thus the nonzero AF order $m$ gives rise to a new term to the original single particle Hamiltonian and spontaneously breaks the TR symmetry of the system. To calculate $\langle\delta n\rangle$, we can diagonalize $\mathcal{H}_0^{(m)}(\bold k)$, which gives four eigenstates $|\chi_{\alpha\beta}(m,\bold k)\rangle=[\mu_{a\uparrow}^{\alpha\beta},\mu_{b\uparrow}^{\alpha\beta}, \mu_{a\downarrow}^{\alpha\beta},\mu_{b\downarrow}^{\alpha\beta}]^T$, with $\alpha,\beta=\pm1$ and the corresponding eigenvalues $E_{\alpha\beta}=\alpha\sqrt{d_1^2+d_5^2+(d_2+\beta d_{15})^2}$. The expectation value of $\langle\delta n\rangle$ is determined by
\begin{eqnarray}\label{eqn:mean1}
\langle\delta n(m,\Delta)\rangle&=&\frac{1}{2N_0}\sum_{\alpha\beta\bold k}f(E_{\alpha\beta})(|\mu_{a\uparrow}^{\alpha\beta}|^2+|\mu_{a\downarrow}^{\alpha\beta}|^2-\nonumber\\
&&-|\mu_{b\uparrow}^{\alpha\beta}|^2-|\mu_{b\downarrow}^{\alpha\beta}|^2).
\end{eqnarray}
where $f(E_{\alpha\beta})$ is the fermi distribution function. Bear in mind these results, we obtain the mean field free energy
\begin{eqnarray}\label{eqn:free1}
F_{HF}(m)&=&\sum_{\alpha\beta\bold k}E_{\alpha\beta}(m,\bold k)f(E_{\alpha\beta})+\frac{N_0}{4}U(1+m^2)\nonumber\\
&&+\frac{N_0}{4}U\langle\delta n(m,\Delta)\rangle^2.
\end{eqnarray}
Combining the eqs. (\ref{eqn:mean1}) and (\ref{eqn:free1}) we can solve the minimum point of the free energy as a function of AF order $m$ at each value of $U$ which increases from zero to some finite number. It is expected when $U>U_{c1}$, the AFMI phase occurs with the value $m>0$ at the minimum point of free energy. The key thing is that once we obtain the AFMI phase at $U>U_{c1}$, we can further determine the topological properties of the system by analyzing the spectrum of the Hamiltonian $\mathcal{H}_0^{(m)}(\bold k)$, in which the nonzero term $d_{15}=mU/2$ results in a spin-dependent renormalization to original single particle bulk gap. The bulk gaps for the spin-up and spin-down states read now ${\cal E}_{gap\uparrow}=|2t_1-|\Delta-mU/2||$ and ${\cal E}_{gap\downarrow}=|2t_1-\Delta-mU/2|$, and the topology of spin-up and spin-down branches will be determined separately. For spin-up branch, the state is topologically nontrivial when $|\Delta-mU/2|<2t_1$ and trivial when $|\Delta-mU/2|>2t_1$, while for spin-down branch, the corresponding conditions become $|\Delta+mU/2|<2t_1$ and $|\Delta+mU/2|>2t_1$, respectively. In the following numerical study we assume the parameters $\Delta,m\geq0$ (other cases of negative parameters can be discussed similarly), in which situation one can see the topology of the spin-down states is more fragile to the interaction than that of the spin-up branch. Intuitively, this is because for spin-up states the interaction compensates the onsite energy difference in the mean field picture while for spin-down states it enhances such difference and blocks the nearest-neighbor hopping.

\begin{figure}[ht]
\includegraphics[width=0.95\columnwidth]{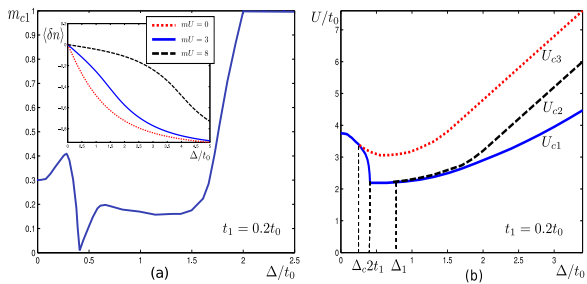}
\caption{(Color online) (a) The mean field value $\langle\delta n(m,\Delta)\rangle$ (inset) and the value of AF order $m_{c1}$ at the transition point $U=U_{c1}$; (b) Relation between $U_{c1}, U_{c2} U_{c3}$ and $\Delta$, where $\Delta_c\simeq0.26t_0$ and $\Delta_1\simeq0.76t_0$. The results are obtained at zero temperature.} \label{free-energy}
\end{figure}
The mean field value $\langle\delta n(m,\Delta)\rangle$ is given in the inset of Fig. \ref{free-energy}, with which one can further determine the spontaneous AF order $m_{c1}$ at the AF phase transition point $U=U_{c1}$, as shown in Fig. \ref{free-energy} (a). For small $\Delta$ ($<2t_1$) where the system is in QSH phase in the non-interacting regime, the value of $m_{c1}$ is finite, and this indicates the phase transition is of the first order. As $\Delta$ increases the first order transition is softened and becomes the second order one around $\Delta=2t_1$, at which point $m_{c1}$ varnishes. Further increasing $\Delta$ the phase transition returns to the first order. Another interesting property is the dependence of the critical interaction $U_{c1}$ on $\Delta$ (Fig. \ref{free-energy} (b), blue line). In the case $\Delta<2t_1$, we find $U_{c1}$ is a monotonously  decreasing function of $\Delta$ and changes rapidly at $\Delta=2t_1$, while for $\Delta>2t_1$ it becomes a monotonously increasing function of $\Delta$. This is because the original single-particle bulk gap decreases to zero when $\Delta$ increases to be $2t_1$, and further increasing $\Delta$ enlarges the single-particle bulk gap. These properties lead to important phenomena for the topological phase transition described below.

\begin{figure}[ht]
\includegraphics[width=0.85\columnwidth]{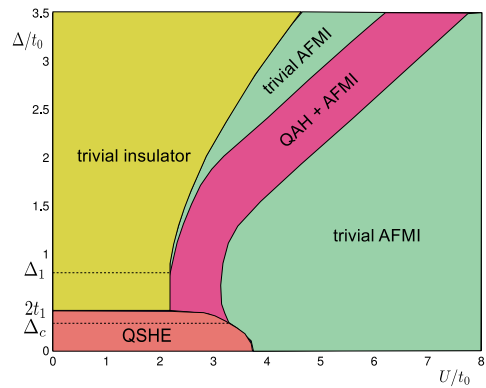}
\caption{(Color online) Mean field phase diagram with the parameter $t_1=0.2t_0$.} \label{phase}
\end{figure}

First, when the on-site energy difference is smaller than some critical value $\Delta<\Delta_c(<2t_1)$, we find $|\Delta+m_{c1}U_{c1}/2|>2t_1$ and $|\Delta-m_{c1}U_{c1}/2|>2t_1$, which indicates once the AF transition occurs both spin branches become topologically trivial. In this case, no gapless edge mode exists at the AFMI phase. Second, when $\Delta_c<\Delta<2t_1$, we have for the spin-down states $|\Delta+m_{c1}U_{c1}/2|>2t_1$, while for spin-up states $|\Delta-m_{c1}U_{c1}/2|<2t_1$. This means for the case $U>U_{c1}$, the spin-down branch becomes trivial but the spin-up branch keeps topologically nontrivial for $U_{c1}<U<U_{c3}$. Here $U_{c3}$ is determined by $m_{c3}U_{c3}/2-\Delta=2t_1$ with $m_{c3}$ obtained at $U=U_{c3}$. In this way, the quantum anomalous Hall (QAH) insulator which has a quantized CHC ($e^2/h$) associated with the AF phase is obtained. Third, when $2t_1<\Delta<\Delta_1$ with $\Delta_1$ depending on the lattice parameters such as hopping coefficients, the system is in the trivial band insulator phase for the weak interacting regime ($U<U_{c1}$). However, it is quite interesting that in the region $U_{c1}<U<U_{c3}$, we find again the phase that the QAH insulator with AF ordering. Finally, for $\Delta>\Delta_1$, increasing the interacting energy to $U>U_{c1}$ first drives the system from trivial band insulator to the non-topological AFMI. However, further increasing $U$ to the region $U_{c2}<U<U_{c3}$ again leads to the topological phase of QAH and AF insulator, with $U_{c2}$ determined via $\Delta-m_{c2}U_{c2}/2=2t_1$. For all these cases, when the interacting strength $U>U_{c3}$, the system turns to the topologically trivial AFMI phase. Fig. \ref{phase} gives the mean field phase diagram in different parameter regimes.

\begin{figure}[ht]
\includegraphics[width=0.8\columnwidth]{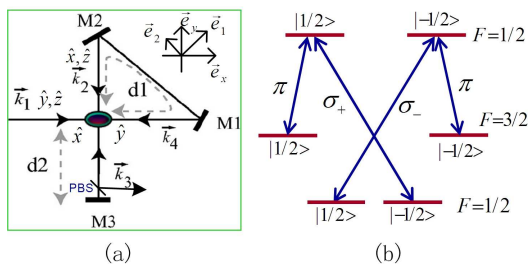}
\caption{(Color online) (a) Cold atoms trapped in a square lattice based on the set-up in Porto's group at NIST \cite{Porto}. (b) Generation of spin-dependent gauge potential with $^6$Li atoms.} \label{configuration}
\end{figure}

The correlation effect obtained here is not restricted in square lattice. For the KM model in the honeycomb lattice, when both the Hubbard-type interaction and sublattice on-site energy difference are present, we have confirmed the same phase diagram given in Fig. \ref{phase}. First-principle calculation shows that the physics in the transition metal oxide Na$_2$IrO$_3$ \cite{Nagaosa} is captured by the KM-like model, and thus this material is a natural candidate to observe the above correlation physics. Alternatively, we propose a novel scheme with cold atomic platform to study these phenomena. The square lattice can be realized by the periodic optical potential $V_{latt}=-V_0(\cos^2k_0x+\cos^2k_0y)-V_1\sin^2[k_0(x+y)/2]$, which is achievable with the experimental platform by Porto's group at NIST (Fig. \ref{configuration}(a)) \cite{Porto} with a minor modification. The second term of the potential (with depth $V_1$) contributes to an anisotropic term, due to which the local $s$-orbital on the $A$ sites extends longer than that of $B$ sites in the $\vec e_1$ direction. This leads to the hopping coefficient $t_1=t_{a1}-t_{b1}>0$. Furthermore, the on-site energy difference between the $s$-orbitals on the $A$ and $B$ sites is controllable and is obtained by $\Delta=V_1+2E_r^{1/2}[(V_{0}-V_1/2)^{1/2}-(V_{0}+V_1/2)^{1/2}]$, with $E_r=\hbar^2 k_0^2/2m$ the recoil energy.

The candidate for the fermi atoms can be $^6$Li and $^{40}$K, etc, and the hyperfine levels of $^6$Li atoms are sketched in Fig. \ref{configuration}(b). The periodic spin-dependent gauge potential can be generated by coupling the atomic hyperfine levels to radiation. In Fig. 4(b) the $\sigma_\pm$ transitions can be realized by a single laser with Rabi-frequency $\Omega_{\sigma}=\hat e_+\Omega_0e^{ik_1(x+y)}+\hat e_-\Omega_0e^{-ik_1(x+y)}$ with $\hat e_\pm$ representing the circular polarizations of photons, while the $\pi$ transitions are achieved by the laser with $\Omega_{\pi}=\hat e_\pi\Omega_1\sin[k_0(x-y)/2+\pi/4]$. For this the $\sigma_+$ and the left hand $\pi$ transitions consist of a $\Lambda$ system, while $\sigma_-$ and the right hand $\pi$ transitions consist of another $\Lambda'$ system. It is known each $\Lambda$ system has a dark-state solution with which we define the (pseudo)spin states by $|\xi_{\uparrow}\rangle=\cos\theta|\frac{1}{2},-\frac{1}{2}\rangle-\sin\theta e^{-i\varphi}|\frac{3}{2},\frac{1}{2}\rangle$ and $|\xi_{\downarrow}\rangle=\cos\theta |\frac{1}{2},\frac{1}{2}\rangle-\sin\theta e^{i\varphi}|\frac{3}{2},-\frac{1}{2}\rangle$ with $\theta$ defined via $\tan\theta=\Omega_1\sin[k_0(x-y)/2+\pi/4]/\Omega_0$ and $\varphi=k_1(x+y)$. The spin-dependent gauge potential in the defined spin-$1/2$ subspace is obtained straightforwardly by \cite{atomgauge1,atomgauge2} $\bold A(\bold r)=\frac{k_1\sin k_0(x-y)}{1+\gamma^2[1+\sin k_0(x-y)]}\vec e_1s_z$ with $\gamma=\Omega_1/\sqrt{2}\Omega_0$, which gives rise to a spin-dependent periodic Peierls phase only for the nearest-neighbor-site hopping, as required in the present model. The interaction strength in cold atoms can be readily manipulated from weak or zero to strong regime, and from attractive to repulsive by Feshbach resonance \cite{cold}.

In conclusion, we have studied the correlation effect on topological insulator in a square lattice. A rich phase diagram is obtained by varying the staggered sublattice potential $\Delta$ and the Hubbard on-site interaction $U$. Especially, for the case $\Delta>\Delta_c$, we see the increasing $U$ can drive the system into the topological antiferromagnetic Mott insulating phase which exhibits the quantum anomalous Hall effect. The observation of these correlation physics in solid state and cold atom systems is discussed. Several interesting issues following the present study deserve future efforts in the research. For example, in square lattice, the frustration appears when the magnitudes of the hopping coefficients $t_1\simeq t_0$. In this case the spin liquid phase with gapless spinon excitations may be obtained in the intermediate interaction strength \cite{Balents}. It is especially interesting to find out in this model the gapless spinon excitation is chiral or helical in different parameter regimes.
%Also, generalization the present 2D model to 3D topological insulators will be another interesting issue to study the possible appearance of nontrivial many-body states when the Mott transition occurs.

X.J.L thanks Congjun Wu, J. Sinova, Ar. Abanov and Chia-Ren Hu for fruitful discussions and helpful comments. This work is supported by NSF under Grant No. DMR-0547875.

%%%%%%%%%%%%%%%%%%%%%%%%%%%%%%%%%%%%%%%%%%%%%

%%%%%%%%%%%%%%%%%%%%%%%%%%%%%%%%%%%%%%%%%%%%%
\noindent


\begin{thebibliography}{99}

%\bibitem{QHE1} K.V. Klitzing, G. Dorda, and M. Pepper,
%Phys. Rev. Lett. {\bf 45}, 494 (1980).

\bibitem{T1} C. L. Kane and E. J. Mele, Phys. Rev. Lett. {\bf 95}, 146802
(2005); $ibid$ {\bf 95}, 226801 (2005).

\bibitem{T1'}B. A. Bernevig et al., Science, {\bf 314}, 1757 (2006).

\bibitem{T1''} J. E. Moore et al., Phys. Rev. B {\bf 75}, 121306(R) (2007).

\bibitem{T2} L. Fu etal.,% C. L. Kane, and E. J. Mele,
Phys. Rev. Lett. {\bf 98}, 106803 (2007); L. Fu and C.L. Kane, \textit{ibid}, {\bf 100}, 096407 (2008).

\bibitem{T3} M. K\"{o}nig et al., Science, {\bf 318}, 766 (2007); D. Hsieh et al., Nature {\bf 452}, 970 (2008);
      D. Hsieh et al., Science {\bf 323}, 919 (2009).

\bibitem{Bernevig} B.A. Bernevig and S.-C. Zhang, Phys. Rev. Lett. {\bf 96}, 106802 (2006).

\bibitem{zhu} S.-L. Zhu etal., Phys. Rev. Lett. {\bf 97}, 240401 (2006);
X. -J. Liu etal.,% X. Liu, L. C. Kwek and C. H. Oh,
\textit{ibid}, {\bf 98}, 026602 (2007);

\bibitem{liu2} X.-J. Liu etal., %X. Liu, L. C. Kwek and C. H. Oh,
Phys. Rev. B \textbf{79}, 165301 (2009).

\bibitem{Levin} M. Levin1 and A. Stern, Phys. Rev. Lett. \textbf{103}, 196803 (2009).

\bibitem{Maciejk} J. Maciejko etal., arXiv: 1004.3628v1.

\bibitem{Haldane} F. D. M. Haldane, Phys. Rev. Lett. {\bf 61}, 2015 (1988).

\bibitem{conductance} A. N. Redlich, Phys. Rev. D, {\bf 29}, 2366 (1984).

\bibitem{liu3} X.-J. Liu, X. Liu, C. Wu, and J. Sinova, Phys. Rev. A {\bf 81}, 033622 (2010).

\bibitem{wu} C. Wu etal., % B. A, Bernevig and S.-C. Zhang,
Phys. Rev. Lett. \textbf{96}, 106401 (2006); C. Xu and J. E. Moore, Phys. Rev. B {\bf 73}, 045322 (2006).

\bibitem{Nagaosa} A. Shitade et al., Phys. Rev. Lett. {\bf 102}, 256403 (2009).

\bibitem{Porto} J. Sebby-Strabley et al., Phys. Rev. A {\bf 73}, 033605 (2006); Phys. Rev. Lett. {\bf 98}, 200405 (2007).

\bibitem{atomgauge1} J. Ruseckas etal., % G. Juzeliunas, P. Ohberg, and M. Fleischhauer,
Phys. Rev. Lett. {\bf 95}, 010404 (2005).

\bibitem{atomgauge2} T. D. Stanescu etal., % C. W. Zhang, and V. Galitski,
Phys. Rev. Lett. {\bf 99}, 110403 (2007); X. -J. Liu etal., \textit{ibid}, \textbf{102}, 046402 (2009).

\bibitem{cold} I. Bloch et al., Rev. Mod. Phys. {\bf 80}, 885 (2008).

\bibitem{Balents} D. Pesin and L. Balents, Nat. Phys. {\bf 6}, 376 (2010).

%\bibitem{experiment1} Y. -J. Lin, R. L. Compton, A. R. Perry, W. D. Phillips, J. V. Porto, and I. B. Spielman,
%Phys. Rev. Lett. {\bf 102}, 130401 (2009).

%\bibitem{hardwall} T. P. Meyrath, F. Schreck, J. L. Hanssen, C.-S. Chuu, and M. G. Raizen,
%Phys. Rev. A {\bf 71}, 041604(R) (2005).

%\bibitem{liu4} X.-J. Liu etal., in preparation.

\end{thebibliography}
\end{document}